УДК 550.347.2 : 550.388.2 : 550.34 : 001.891.57. = 20

# Ionospheric total electron content variations observed before earthquakes: Possible physical mechanism and modeling

## A.A. Namgaladze[1], O.V. Zolotov[1], I.E. Zakharenkova[2], I.I. Shagimuratov[2], O.V. Martynenko[1]

[1] *Polytechnical Faculty of MSTU, Physics Chair, Murmansk*
[2] *West Department of IZMIRAN, Kaliningrad*

**Аннотация.** На основе GPS наблюдений полного электронного содержания (ТЕС) ионосферы с использованием глобальных и региональных карт ТЕС, а также отдельных спутниковых измерений, обнаружены возмущения ТЕС, предшествующие сильным землетрясениям. Для сильных среднеширотных землетрясений сейсмо-ионосферные аномалии выглядят как области локального увеличения или уменьшения ТЕС в окрестности эпицентра готовящегося землетрясения. Такие структуры формируются в ионосфере за несколько дней до основного сейсмического события. Величина наблюдаемых модификаций ионосферной плазмы достигает 30-90 % по сравнению с невозмущенными фоновыми значениями. Область максимального проявления аномалии имеет пространственные размеры более 1500 км по широте и 3500-4000 км по долготе. В случае сильных низкоширотных землетрясений обнаруживаются эффекты, связанные с модификацией экваториальной аномалии: углубление или "заполнение" ионосферного провала электронной концентрации на магнитном экваторе.

Предложен физический механизм формирования указанных аномалий. Мы полагаем, что наиболее вероятной причиной наблюдавшихся перед землетрясениями возмущений NmF2 и ТЕС является вертикальный дрейф ионосферной плазмы в F2-области под воздействием зонального электрического поля сейсмического происхождения. Для проверки этой гипотезы проведен ряд модельных расчётов при помощи глобальной самосогласованной численной модели верхней атмосферы Земли UAM (Upper Atmosphere Model). Предложено пространственное распределение электрического потенциала на границе эпицентральной области, необходимое для существования указанного электрического поля. Параметры верхней атмосферы, предшествующие готовящемуся сильному землетрясению, моделировались путём включения дополнительных источников электрического поля в уравнение для электрического потенциала модели UAM, которое решалось численно совместно со всеми остальными уравнениями модели (непрерывности, движения и теплового баланса) для нейтрального и ионизованных газов. Эффективность предложенного механизма исследовалась путём численного моделирования отклика ионосферы на воздействие зонального электрического поля, порождённого сейсмогенными источниками для случаев средних и низких широт. Представлены результаты соответствующих модельных расчётов электрического поля и генерированных им эффектов в F2-области ионосферы и плазмосфере Земли. Результаты моделирования хорошо согласуются с данными наблюдений ТЕС перед сильными землетрясениями как в случае средних, так и низких широт, как по пространственному масштабу, так и по магнитуде наблюдаемых аномалий.

**Abstract.** The GPS derived TEC disturbances before earthquakes were discovered in the last years using global and regional TEC maps, TEC measurements over individual stations as well as measurements along individual GPS satellite passes. For strong mid-latitudinal earthquakes the seismo-ionospheric anomalies look like local TEC enhancements or decreases located in the vicinity of the forthcoming earthquake epicenter. Such structures are generated in the ionosphere for several days prior to the main shock. The amplitude of plasma modification reaches the value of 30-90 % relative to the non-disturbed level. The zone of the anomaly maximum manifestation extends larger than 1500 km in latitude and 3500-4000 km in longitude. In case of strong low-latitudinal earthquakes there are effects related with the modification of the equatorial F2-region anomaly: deepening or filling of the ionospheric electron density trough over the magnetic equator.

The possible physical mechanism which can cause such anomalies has been proposed. We consider that the most probable reason of the NmF2 and TEC disturbances observed before the earthquakes is the vertical drift of the F2-region ionospheric plasma under the influence of the zonal electric field of seismic origin. To check this hypothesis, the model calculations have been carried out with the use of the UAM (Upper Atmosphere Model) – the global numerical model of the Earth's upper atmosphere. The electric potential distribution at the near-epicenter region boundary required for the electric field maintenance has been proposed. The upper atmosphere state, presumably foregone a strong earthquake, has been modeled by means of switching-on of additional sources of the electric field in the UAM electric potential equation which was solved numerically jointly with all other UAM equations (continuity, momentum and heat balance) for neutral and ionized gases. The efficiency of the proposed mechanism has been investigated by means of model calculations of the ionosphere response to the action of zonal electric field produced by seismogenic sources located at the middle and low latitudes. The results of the corresponding numerical model calculations of the electric field and its effects in the ionospheric F2-layer and plasmasphere have been presented. They have revealed a fine agreement with TEC anomalies observed before strong earthquakes at the middle and low latitudes both in spatial scales and in amplitude characteristics.

**Ключевые слова:** ионосферные предвестники землетрясений, эпицентр, сейсмогенное электрическое поле, полное электронное содержание

**Key words:** ionospheric earthquake precursors, epicenter, seismogenic electric field, total electron content

## 1. Introduction

Searches of the seismo-ionospheric precursors have been intensively conducted during last 2-3 decades on the basis of different ground-based and satellite observations, also by means of the special satellite projects of the natural hazards monitoring ["COMPASS-1", "COMPASS-2", "Sich-1M", "QuakeSat", "DEMETER"] (see reference list in (*Pulinets, Boyarchuk*, 2004)). Nowadays there are vast possibilities to investigate the ionosphere modifications and in particular the ionospheric effects associated with seismic activities by use of the satellite Global Positioning System (GPS) signals measurements (*Liu et al.*, 2002; *Plotkin*, 2003; *Liu et al.*, 2004; *Afraimovich et al.*, 2004; *Krankowski et al.*, 2006; *Zakharenkova et al.*, 2007a). The dense network of GPS receivers (a few thousands all over the world) fulfills simultaneous coverage in global scale with high temporal resolution. GPS technique provides measurements of the group and phase delays of the signals $L_1$=1575 MHz and $L_2$=1228 MHz with a 30-sec interval. The ionospheric delay can be transformed into the content of electrons along the signal path between a GPS satellite and GPS receiver, and then recalculated into its vertical projection. The vertical total electron content (TEC) is very sensitive to changes of the maximal electron concentration (NmF2) in the F2 layer of the ionosphere.

The extensive studies of the ionospheric earthquake precursors in the GPS TEC measurements carried out in the last years have revealed that for strong mid-latitudinal earthquakes the seismo-ionospheric anomalies very often (usually) look as local TEC increases and they are situated in the immediate vicinity of the earthquake epicenter area. The zone of the anomaly maximum manifestation (TEC enhancement more than 35 %) has a spatial scale of some thousands km in longitude and about 1000 km in latitude (*Zakharenkova et al.*, 2006; 2007a; 2007b). These sizes are in a good agreement with results obtained from the combined analysis of ground-based and satellite ionosonde measurements. It was found out that the spatial scale of the seismo-modified area of the ionosphere at the heights of the F-layer maximum during the strong earthquake preparation time has the diameter of 20°-40° in geographical coordinates (*Pulinets*, 1998; *Strakhov, Liperovsky*, 1999). The size of the area changes with the earthquake magnitude. *Pulinets et al.* (2005) analyzed the GPS TEC measurements for the strong Mexico earthquake and found out an anomalous ionospheric modification in the spatial (latitude-longitude) distribution of the TEC deviation. The anomalous enhancement of the TEC was registered 3 days prior to the event and it reached the value of 55 % relative to the background conditions.

The negative TEC disturbances (TEC decreases) were also observed, for example for strong Turkey earthquakes of 1999 (*Ruzhin et al.*, 2002) and for several Taiwan and Sumatra earthquakes (*Liu et al.*, 2006; *Liu, Chen*, 2007).

Fig. 1 presents characteristic TEC anomalies observed prior to the mid-latitudinal earthquakes of 25 September 2003, Japan and 8 January 2006, Greece (*Zakharenkova et al.*, 2007a; 2007b).

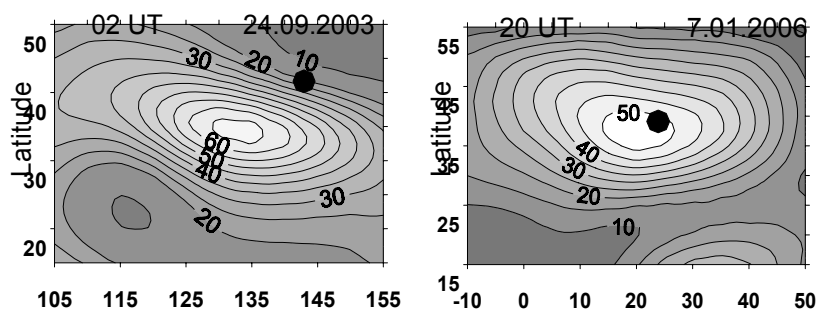

Fig. 1. Differential TEC (%) maps 1 day prior to: 1) the Japanese earthquake of September 25, 2003 (M=8.3), 2) the Greece earthquake of January 8, 2006 (M=6.8). The epicenter position is marked by the black dot

In case of strong low-latitudinal earthquakes there are effects related with the modification of the ionospheric F2-region equatorial anomaly: increase or decrease of the equatorial anomaly with trough deepening or filling (*Depueva, Ruzhin*, 1995; *Depueva, Rotanova*, 2001; *Pulinets, Legen'ka*, 2002; *Depueva et al.*, 2007). Fig. 2 presents the seismogenic equatorial anomaly observed in the measurements of Alouette-2 satellite 1 day prior to the Chile earthquake of 12 April 1963 (*Depueva, Ruzhin*, 1995). It was found out that one day before the event the latitudinal dependence of the critical frequency (foF2) on the magnetic inclination (I) looked like a curve with two maxima symmetrically located around the magnetic equator in midnight hours of the local time. The plasma concentration over the epicentral area was reduced more than 10 times related to the normal conditions.

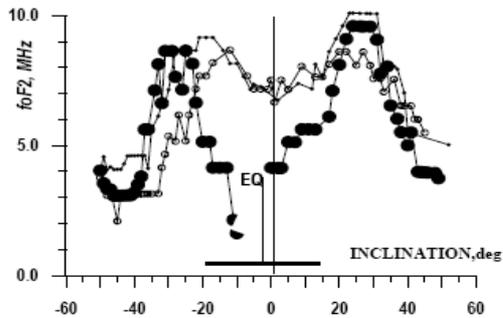

Fig. 2. Seismogenic equatorial anomaly (black circles) in the ionosphere 1 day prior to the Chile earthquake of 12 April 1963 (Alouette-2 measurements). Other lines – foF2 variations in non-disturbed time (before and after the earthquake). EQ – the epicenter position (*Depueva, Ruzhin*, 1995)

*Liu et al.* (2001) analyzed the measurements of the Taiwan GPS network and reported significant decreases in the TEC on the 3rd and 4th days before the Chi-Chi earthquake. *Liu et al.* (2004) further examined the GPS TEC during all of the 20 M ≥ 6.0 earthquakes at the Taiwan area from September 1999 to December 2002. They found that anomalous decreases in the GPS TEC often appeared 1-5 days before the earthquakes. These effects were probably related with the equatorward displacement of the equatorial anomaly northern crest and reduction of the electron concentration in the crest.

## 2. Physical mechanisms of the seismo-ionospheric precursors appearance

Physical interpretation of the seismo-ionospheric precursors appearance has been proposed in numerous papers (*Sorokin, Chmyrev*, 1999; *Pulinets, Boyarchuk*, 2004) and it is mainly based on the hypothesis about the seismogenic electric field related with the vertical turbulent transportation of the injected aerosols and radioactive particles (radon isotopes). The increase of the atmospheric radioactivity level during the earthquake preparation leads to the enlargement of the ionization and electric conductivity of the near-ground atmosphere. The joint action of these processes leads to the intensification of the electric field in the ionosphere up to the value of units-tens mV/m (*Chmyrev et al.*, 1989).

There are strong arguments in favour of the hypothesis about the seismogenic electric field: a) geomagnetic conjugation of the ionospheric precursors (*Pulinets et al.*, 2003), b) effects related with the ionospheric F2-region equatorial anomaly controlled by the zonal electric field (*Depueva, Ruzhin*, 1995; *Pulinets, Legen'ka*, 2002).

It was proposed (*Namgaladze et al.*, 2007) that the main reason of the appearance of the local anomalies in the form of the increased (decreased) total electron content of the ionosphere, observed on the base of GPS signals measurements, is the vertical drift of F2-region ionospheric plasma upward (downward) under the influence of the zonal electric field of seismogenic origin directed to the east (west). In the middle latitudes the vertical upward component of the electromagnetic drift, created by the eastward electric field, leads to the increase of the F2 region electron concentration maximum (NmF2) and TEC due to the plasma transportation to the regions with lower concentration of the neutral molecules $O_2$ and $N_2$ and, consequently, with lower loss rate of dominating ions $O^+$ in the ion-molecular reactions (*Brunelli, Namgaladze*, 1988).

The electric field of the opposite direction (westward) creates the opposite – negative – effect in NmF2 and TEC. In the low latitude regions (near the geomagnetic equator) the increase of the eastward electric field leads to the deepening of the equatorial anomaly minimum ("trough" over the magnetic equator in the latitudinal distribution of electron concentration) due to the intensification of the fountain-effect.

## 3. Model calculations

To check this hypothesis on the zonal electric field as the most probable cause of the observed TEC disturbances before earthquakes, the model calculations were carried out with the use of the UAM – the global numerical model of the Earth's upper atmosphere (*Namgaladze et al.*, 1988; 1991; 1998). This first principle model describes the Earth's upper atmosphere behavior by means of solving a system of coupled time-dependent 3D continuity, momentum and heat balance equations for the neutral and ionized atmospheric and ionospheric gas components as well as the equation for the electric potential. The model covers the height range from 60 km up to 100 000 km and takes into account the offset between the geographic and geomagnetic axes of the Earth.

The upper atmosphere state, presumably foregone a strong earthquake, was modeled by means of switching-on of additional sources of the electric field in the UAM electric potential equation which was solved numerically jointly with all other UAM equations (continuity, momentum and heat balance) for neutral and ionized gases.

These sources were switched on and maintained as permanent during 24 h in the form of additional positive and negative potentials with values of 2 and 5 kV (in case of the low-latitudinal sources) and of 10 kV (in case of the mid-latitudinal source) on the western and eastern boundaries of near-epicentral areas consequently.

In this analysis we have investigated 2 near-epicentral areas with sizes of 10° in latitude and 30° in longitude; epicenters have been situated at the points with magnetic coordinates: 1) 45N, 90; 2) 15S, 210. Such sizes approximately correspond to the horizontal sizes of the regions of the TEC increased values which were found out in (*Pulinets et al.*, 2003; *Zakharenkova et al.*, 2007b). The first region is a typical middle-latitude ionosphere, the second region is a near-equatorial ionosphere, in which the effects of the electric field are more essential than at the middle latitudes.

In Fig. 3 the numerical grid of geomagnetic coordinates used in the model calculations is shown and its mesh points in which additional potentials were set, are marked by the circles (dark circles correspond to the positive potential, light circles correspond to the negative potential).

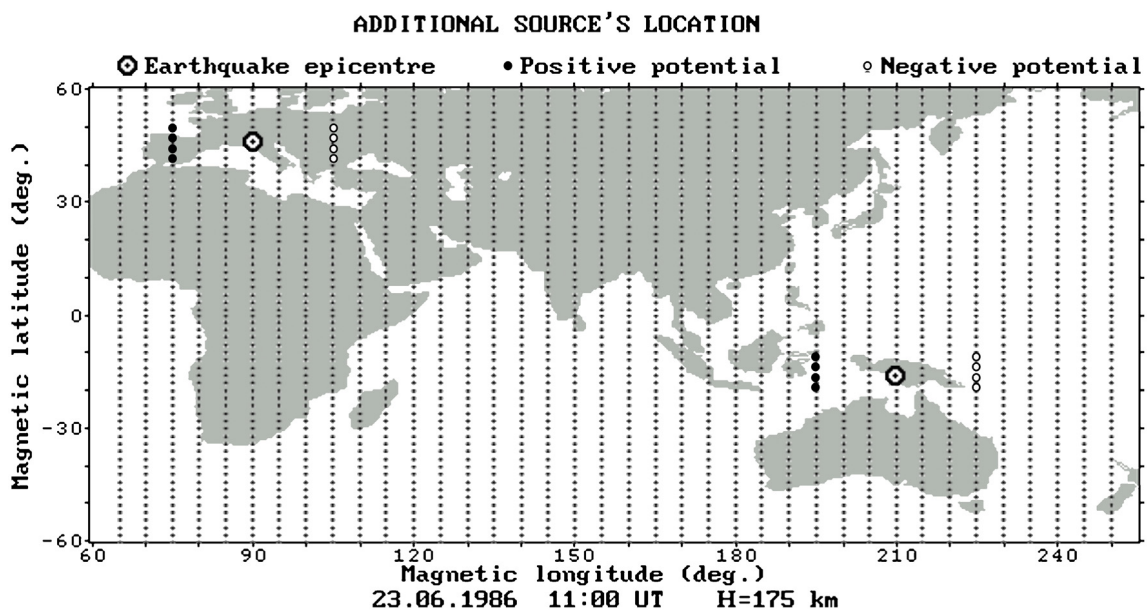

Fig. 3. The numerical grid of geomagnetic coordinates used in the model calculations and its mesh points in which the additional potentials were set noted by the circles (dark circles correspond to the positive potential, light circles correspond to the negative potential)

**4. Ionospheric effects created by additional sources of the electric field**

The analysis of results of model calculations was carried out on the basis of comparison of the global maps of various ionospheric parameters obtained in quiet (without an additional electric field) and disturbed (with additional – presumably seismogenic – sources of an electric field) conditions. For the quiet condition the magneto-quiet day of a June solstice at high solar activity was accepted.

The calculated electric potential pattern and horizontal electric field vectors for the sources of 5 and 10 kV per mesh point are shown in Fig. 4 for the quiet and disturbed conditions. From this Figure the regions of eastward electric field above the prospective epicenters of the future earthquakes, appeared when imposing (switching on) additional sources, and the similar magneto-conjugated to them regions in the opposite hemisphere are visible. The symmetry of the potential and electric field concerning the geomagnetic equator is caused by the ideal electroconductivity of the ionospheric plasma along the geomagnetic field lines and, accordingly, by their electric equipotentiality. The amplitudes of the additional eastward electric fields are of about 2-4 mV/m in case of the low-latitude source and 4-10 mV/m in case of the mid-latitude source. They exceed the background quiet fields (of about 0.2 and 1 mV/m, correspondingly), but they are noticeably smaller than the quiet high-latitude electric fields of magnetospheric origin (15-25 mV/m) obtained in the model calculations.

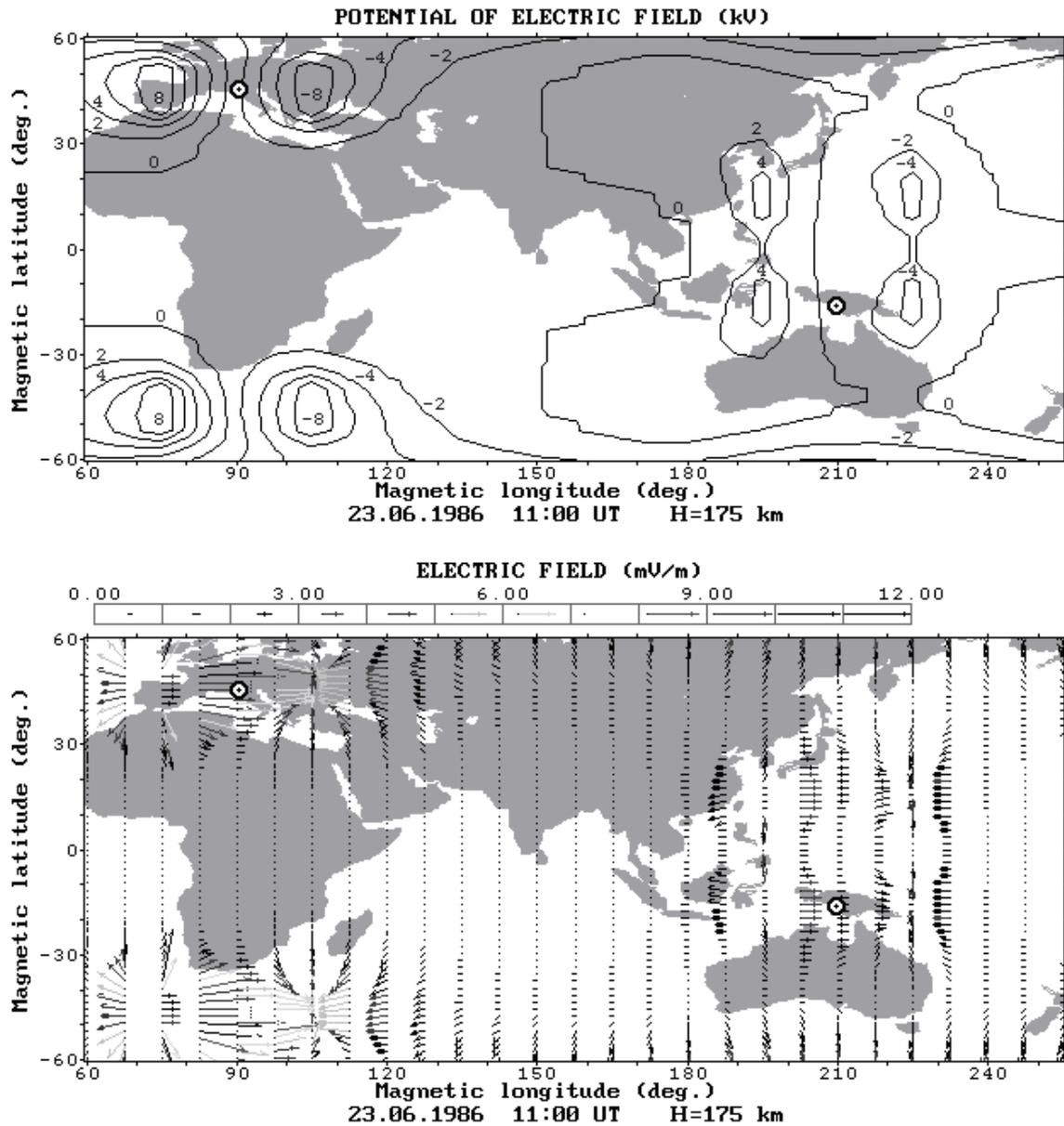

Fig. 4. Model calculated electric field potential (top) and horizontal electric field vectors (bottom) generated by the background and additional (seismogenic) sources for two earthquake epicenters indicated by the markers

The calculated ionospheric effects in TEC and foF2 (F2-layer critical frequency) created by additional sources of the electric field are shown in Figs. 5 and 6. As we can see from these Figures, the action of the near-equatorial source intensifies the equatorial anomaly of the F2 layer in the near-epicentral area of the ionosphere by deepening the minimum of foF2 over the magnetic equator and displacing the anomaly crests from the equator to the middle latitudes. This behavior agrees completely with the Alouette-1 and 2 observations (*Depueva, Ruzhin*, 1995; *Ruzhin, Depueva*, 1996; *Depueva et al.*, 2007) as it is seen from the comparison of Figs. 2 and 5, 6.

The action of the mid-latitudinal source leads to the increase of foF2 and TEC in the near-epicentral and magnetically conjugated areas of the ionosphere if these regions are lighted by the Sun as our calculations show. Inclusion of the source at night does not cause appreciable effects, they appear after sunrise and are kept after sunset. The amplitude of enhancements and its spatial sizes are in a good agreement with the corresponding characteristics of precursors in the TEC observations. Although the electric field magnitudes in case of the mid-latitudinal source in our calculations are 2-5 times more than in case of low-latitudinal ones, the evoked effects are weaker than in the near-equatorial regions.

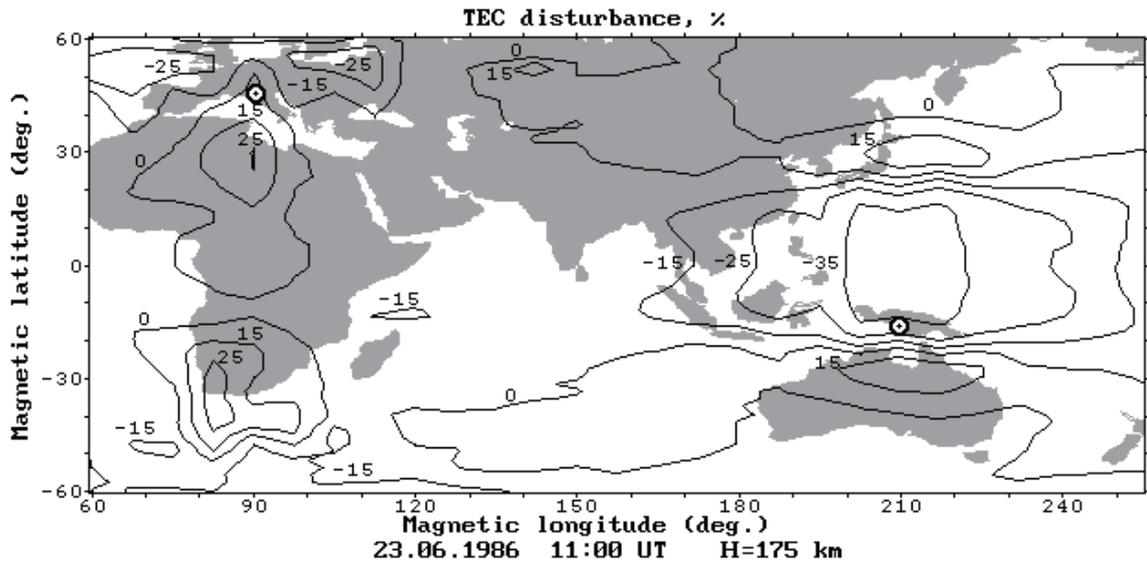

Fig. 5. Model calculated relative TEC disturbances (%) induced by the additional (seismogenic) sources for two earthquake epicenters indicated by the markers

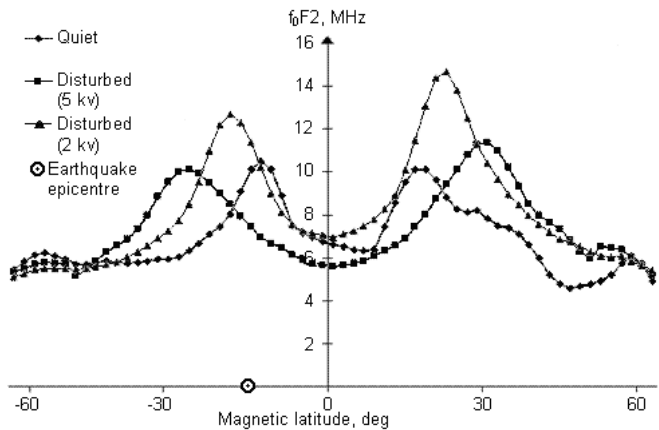

Fig. 6. Model calculated latitudinal variations of the F2 layer critical frequencies along the magnetic meridian of the equatorial earthquake epicenter located at 15° magnetic latitude for quiet and disturbed conditions

**5. Summary**

As our model calculations show, the very probable reason of the NmF2 and TEC disturbances observed before earthquakes is the vertical drift of the F2-region ionospheric plasma under the influence of the zonal electric field of seismogenic origin. In case of TEC enhancements in the middle latitudes and deepening of the equatorial anomaly trough this field is directed to the east and induces the electromagnetic drift of the plasma across the geomagnetic field with velocity directed straight upwards over the magnetic equator and upwards and pole wards in the middle latitudes.

The upward plasma drift in the middle latitudes provokes the increase of electron concentration in the F2 region of the ionosphere due to decrease of the dominating ions $O^+$ loss rate in the ion-molecular reactions; in the low latitudes it provokes decrease of NmF2 over the magnetic equator due to the fountain-effect intensification.

The pattern of the spatial distribution of the seismogenic origin electric field potential has been proposed. For existence of the eastward electric field in the near-epicentral area it is necessary to dispose the positive electric charges on the western boundary of this area and the negative charges on the eastern boundary.

The efficiency of the proposed mechanism has been investigated by means of the numerical model calculations of the ionosphere response to the action of the zonal electric field with amplitude values of the order of 2-10 mV/m presumably produced by seismogenic sources located in the middle and low latitudes. Results of the model calculations have revealed the fine agreement with the TEC and NmF2 anomalies observed before strong earthquakes in the middle and low latitudes both in spatial scales and in amplitude characteristics. We do not discuss here the mechanism of generation of the seismogenic electric field and we do not explain, why the positive charges should accumulate on the western boundary of the near-epicentral region, and negative charges – on the eastern boundary or vice versa. We show only what kind of electric field can produce the TEC and NmF2 disturbances observed before strong earthquakes.

The work was partially supported by the Russian Foundation for Basic Research, grant No. 08-05-98830.